\documentclass[a4paper,10pt,twocolumn]{article}
\pdfoutput=1

\usepackage{mathpazo}
\usepackage{amsmath}

\usepackage[utf8]{inputenc}

\usepackage[american]{babel}
\usepackage[font=small,labelfont=bf]{caption}

\usepackage[round]{natbib}
\usepackage{url}

\usepackage{graphicx}
\graphicspath{{./}{arxivfigs/}}

\DeclareMathOperator{\var}{var}

\DeclareMathOperator{\hilbert}{H}
\DeclareMathOperator{\prob}{Pr}
\newcommand{\dd}{\mathrm{d}}    
\newcommand{\ii}{\mathrm{i}}    
\newcommand{\lpi}{\pi}          
\renewcommand{\varpi}{\pi}
\renewcommand{\vec}[1]{\boldsymbol{#1}}
\newcommand{\op}[1]{\boldsymbol{#1}}

\begin{document}

\title{\bf Mental States as Macrostates\\Emerging from EEG Dynamics}
\author{Carsten Allefeld \and Harald Atmanspacher \and Ji\v rí Wackermann}
\date{\normalsize Institute for Frontier Areas of Psychology and Mental Health, Freiburg, Germany\\[1em]
Preprint, final version published as:\\
Mental states as macrostates emerging from brain electrical dynamics.\\
\emph{Chaos}, 19(1):015102, 2009. doi:10.1063/1.3072788}

\maketitle

\begin{abstract}
Correlations between psychological and physiological phenomena form the basis for different medical and scientific disciplines, but the nature of this relation has not yet been fully understood. One conceptual option is to understand the mental as ``emerging'' from neural processes in the specific sense that psychology and physiology provide two different descriptions of the same system. Stating these descriptions in terms of coarser- and finer-grained system states (macro- and microstates), the two descriptions may be equally adequate if the coarse-graining preserves the possibility to obtain a dynamical rule for the system. To test the empirical viability of our approach, we describe an algorithm to obtain a specific form of such a coarse-graining from data, and illustrate its operation using a simulated dynamical system. We then apply the method to an electroencephalographic (EEG) recording, where we are able to identify macrostates from the physiological data that correspond to mental states of the subject.
\end{abstract}

\section{Introduction}

The existence of correlations between psychological and physiological phenomena, especially brain processes, is the basic empirical fact of psychophysiological research. Relations between mental processes, including modes of consciousness, and those occurring in its physical ``substrate'', the central nervous system, are generally taken as a matter of course: They form the basis for the use of drugs in the treatment of mental disorders in psychiatry, they are applied as a research tool to shed light on the details of psychological mechanisms, and the explication of the neural structures underlying mental functioning forms the subject of cognitive neuroscience. Still, it remains unclear what the nature of the observed correlations is and what exactly is to be conceived as a neural correlate of a psychological phenomenon.%
\footnote{While the recent discussion focuses on the notion of ``neural correlates of consciousness'' \citep[cf.][]{metzinger:neural}, we are interested in psychophysiological correlations in general.}

One way to approach these issues is to interpret the mental as a domain \emph{emerging} from an underlying physiological domain \citep{broad:mind, beckermann:emergence}. However, despite its long history reaching to recent scientific contributions \citep[e.g.][]{darley:emergent, seth:measuring}, the term emergence is not very well defined and it is used in a large number of different meanings \citep[cf.][]{stephan:emergentism, oconnor:emergent}.

In our understanding, emergence is a relation between different \emph{descriptions} of the same system. In this view, the occurrence and correlation of psychological and physiological phenomena is due to the fact that the object of psychophysiological research (the research subject) can be approached and examined in different ways. More specifically, emergence is to be conceived as a relation between different descriptions each of which is useful or adequate in its own manner. The question arises how there can be more than one adequate description for the same system, and what has to be the nature of their relation in order to permit this.%
\footnote{Note that we are concerned with the atemporal or ``synchronous'' structure of such a relation between descriptions, and do not address the question of how a phenomenon emerges ``diachronically'', in a process unfolding in time.}

In this paper we present one possible answer to these questions, motivated by ideas on the emergence of mental states from neurodynamics introduced by \citet{atmanspacher:contextual}, where the two descriptions take on the form of a dynamical system. We introduce the further specification that the relation between the two associated state spaces is characterized by a Markov coarse-graining (Sec.~\ref{conceptual}), which leads us to consider metastable states as a particular form of emergent states. In order to demonstrate the practical viability of these ideas, we develop a method to identify metastable states from empirical data (Sec.~\ref{algorithm}), and illustrate the operation of the algorithm using data from a simulated system (Sec.~\ref{example}). In the application of the method to a recording of brain electrical activity, we are able to identify states closely corresponding to the mental states of a subject, based on the analysis of the EEG data alone (Sec.~\ref{application}).

\section{Emergence in dynamical systems}
\label{conceptual}

A descriptive approach that has proven very fruitful in physics and other fields of the natural sciences is utilizing the concept of a \emph{dynamical system} \citep{robinson:dynamical, chan:chaos}. Such a description is formulated with respect to the \emph{states} the system can assume, and a dynamical rule that defines the way the state of the system evolves over time. The possible system states form a \emph{state space}, which in the most general case is just a set of identifiable and mutually distinguishable elements.%
\footnote{This is in accordance with the concept of system states in cybernetics and related disciplines \citep[cf.][]{ashby:principles}, but is at variance with the use of the term in physics where a state space is generally taken to be spanned by a set of observables (properties that can be precisely quantified). Such a less structured concept of state space is useful because it also covers cases where it is not obvious how to endow that space with a formal structure, for instance mental states. However, as \citet{gaveau:dynamical} point out, introducing into a state space a dynamics in the form of transition probabilities (see below) implicitly provides it with a metric structure.}

For a well-defined relation between two such descriptions to hold, it is necessary that the two state spaces can be related to each other.%
\footnote{This of course does not have to be the case; different descriptions of the same system may also be incompatible with each other.}
A simple possibility is that the system assumes a particular state in one description exactly if it is in any out of a certain set of states of the other description; that is to say, one state space is a \emph{coarse-graining} or \emph{partition} of the other state space. Because of this asymmetry between the two descriptions one may speak of a higher-level and a lower-level description, and refer correspondingly to \emph{macrostates} and \emph{microstates} of the system. The classic example in physics for this kind of inter-level relation is that between the phenomenological theory of thermodynamics, dealing with the macrostates of extended systems defined in terms of observables such as temperature and pressure, and the theory of statistical mechanics, relating them to microstates defined in terms of the constituents of those systems.%
\footnote{In this context the terms macrostate and microstate derive from the circumstance that they refer to the properties of a ``macroscopic'' system versus those of its ``microscopic'' constituents. Though these terms often imply a difference in spatiotemporal scale, the important point is the difference in the amount of detail given by the descriptions.}

The description of a system is chosen by an observer, but it is also subject to objective constraints insofar as different descriptions may be differently adequate or useful. For a description as a dynamical system, the adequacy of a particular set of system states becomes apparent in the possibility to find a \emph{dynamical rule}, $\Phi_{\Delta t}$, whereby the current state $x_t$ of the system determines its further evolution,
$$
x_{t + \Delta t} = \Phi_{\Delta t} \left ( x_t \right);
$$
here $t$ is a continuous or discrete time variable and $\Delta t$ a time interval. A particular state space definition may therefore be called \emph{dynamically adequate} if the specification of a state implies all the available information which is relevant for determining subsequent states, that is, if in this description the system possesses the \emph{Markov property}  \citep[cf.][]{shalizi:macrostate}. In this sense, the most general model of a dynamical system is the Markov process---a stochastic model which includes deterministic dynamics as a limiting case \citep[cf.][]{chan:chaos}.

It is important to note that this criterion for selecting a descriptive level implies a reference back to that same level; while employing a more fine-grained set of states may serve to improve the prediction of the future of a system in general, it will in most cases result in a loss of the Markov property with respect to these finer-grained states themselves. In other words, the Markov-property criterion distinguishes descriptive levels at which the system exhibits a self-contained dynamics (``eigendynamics''), independent of details present at other levels.%
\footnote{This concept is akin to the idea of operational closure or autonomy in the theory of autopoietic systems \citep{maturana:autopoiesis, varela:principles}, which alongside the separation from the environment also refers to the indifference of system operations towards the internal ``microscopic'' complexity of system elements \citep[cf.][]{luhmann:social}. However, the topic of self-defined system boundaries is not addressed in this paper and accordingly, the term ``system'' is used in the unspecific sense of a section of reality which has been chosen for observation.}

This specification of the kind of descriptions sought for leads to a more specific concept of emergence as an inter-level relation.%
\footnote{The stability conditions of \citet{atmanspacher:contextual} are here realized by the Markov property, while contextual constraints \citep{bishop:contextual} can be seen effective in the selection of a particular descriptive level out of those admissible.}
Given a microscopic state description exhibiting the Markov property, an adequate higher-level description or coarse-graining should ideally preserve it. In the context of deterministic nonlinear systems, where the dynamics is defined by a map from a metric space onto itself, such a coarse-graining is called a Markov partition \citep{adler:symbolic, bollt:markov}; for the general case of stochastic dynamics we propose the term \emph{Markov coarse-graining} \citep[cf.][]{gaveau:dynamical}. Accordingly, states of a higher-level description may be called dynamically emergent states if they correspond to a Markov coarse-graining of a lower-level dynamics.

Interpreting psychophysiological correlations as reflecting a relation of emergence between two levels of description as a dynamical system, the lower-level description is stated in terms of physiological, neural states, the higher-level description in terms of mental states. At both levels a wide variety of descriptive approaches is possible, depending on the experimental methods used to assess the brain state on the one hand (electrophysiology, imaging methods, brain chemistry, etc.) and the chosen set of psychological categories on the other hand (conscious/unconscious, sleep stages, moods, cognitive modes, etc.).%
\footnote{Since each mental state allows for multiple realizations at the neural level, mental states may be said to ``supervene on'' brain states \citep[cf.][]{kim:supervenience}---but this alone does not provide a sufficient characterization of their relation. Moreover, contrary to assumptions prevalent in the discussion \citep[cf.][]{chalmers:what} a neural correlates need not necessarily be realized in a particular neural subsystem of the brain.}
Applying the dynamical specification of emergence outlined above, emergent macrostates that are defined via a Markov coarse-graining of the neural microstate dynamics are candidates for a further characterization as mental states. In order to empirically substantiate these ideas, macrostates obtained from the dynamics that has been observed in neurophysiological data are to be related to mental states of subjects that have been determined by other means, such as behaviorial assessment or verbal reports.

In the following we undertake first steps towards this program. Since a general algorithm for finding Markov coarse-grainings is not known, we focus on the special case of \emph{metastable states}. Because a system stays in such a state for prolonged periods of time and only occasionally switches into another, antecedent states provide practically no information on the subsequent evolution beyond that implied in the current state, so that the macrostate dynamics is approximately Markovian. The following section describes an algorithm to obtain metastable states from the microstate dynamics observed in empirical data.

\section{Identifying metastable macrostates from data}
\label{algorithm}

Metastable states correspond to the ``almost invariant sets'' of a dynamical system, i.e. subsets of the state space which are approximately invariant under the system's dynamics. Since we are dealing with empirical data where there is generally no precise theoretical knowledge of the dynamics, it has to be determined from the data.

Via a finite set of microstates resulting from a discretization of the state space (Sec.~\ref{alg:disc}), the time evolution operator $\Phi_{\Delta t}$ is estimated in the form of a matrix of transition probabilities $\op P$ (Sec.~\ref{alg:dyn}). Metastable states are then determined using an algorithm to find the almost invariant sets of a Markov process (Sec.~\ref{alg:almost}). Additionally, an estimate of the optimal number of macrostates is obtained via an analysis of the characteristic timescales of the dynamics (Sec.~\ref{alg:time}). Our algorithm builds on work by \citet{deuflhard:robust}, \citet{gaveau:dynamical}, and \citet{froyland:statistically}, and re-uses an idea of \citet{allefeld:detecting}.

\subsection{Discretization of the microstate space}
\label{alg:disc}

In order to represent the observed microstate dynamics as a finite-state Markov process, the state space defined by $K$ variables $(x_1, x_2, \ldots, x_K) = \vec x$ has to be discretized, resulting in a set of compound microstates which forms the basis for further analysis. Since the data set may be high-dimensional and of varying density in different areas of the state space, we need a flexible algorithm which adapts the size and shape of microstate cells to local properties of the distribution of data points.

This procedure has to meet two competing goals: It should capture as much detail as possible in order to faithfully represent the underlying continuous dynamics within its discretized version; but since transition probabilities between cells are to be estimated, the number of data points per cell should not fall below a certain minimum. Moreover, the extensions of the cells in the directions of the different variables should be of roughly the same size.%
\footnote{We assume at this point that the variables spanning the state space permit a comparison of distances along different directions. Where this is not the case it is advisable to map all variables onto the same range of values before performing the discretization.}

\begin{figure}
\centering \includegraphics{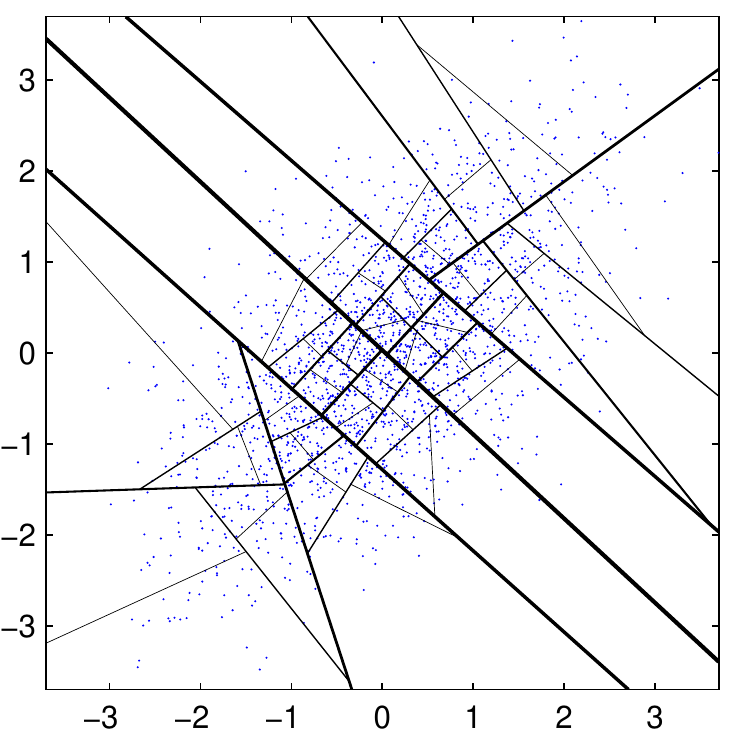}
\caption{Discretization by recursive bipartitioning, illustrated with a set of data points drawn from a two-dimensional normal distribution stretched out along the main diagonal. Cuts occurring earlier in the procedure are indicated by thicker lines.}
\label{bipart}
\end{figure}

To achieve this, we use a recursive bipartitioning approach (Fig.~\ref{bipart}): For a given set of $n$ data points $S = \{\vec x_m\}$, $m = 1 \ldots n$, the direction of maximal variance is determined, i.e. a unit vector $\vec e$, $|\vec e| = 1$, such that $\var_m \left ( \vec x_m \cdot \vec e \right )$ obtains its maximum value. Using the median $M$ of the data points' positions along this direction as a threshold value, the set is divided into two subsets,
\begin{eqnarray*}
S_1 &=& \{ \vec x_m ~|~ \vec x_m \cdot \vec e \leq M \}, \\
S_2 &=& \{ \vec x_m ~|~ \vec x_m \cdot \vec e > M \}.
\end{eqnarray*}
The procedure is repeated for each of the resulting subsets, up to a recursion depth of $b$ steps. This algorithm leads to a practically identical number of data points per cell (either $\lfloor n / 2^b \rfloor$ or $\lceil n / 2^b \rceil$) which can be adjusted via the parameter $b$. It provides a high level of detail in those areas of the state space where the system spends most of the time, and it avoids too elongated cells by applying cuts perpendicular to the current main extension.

\subsection{Microstate dynamics}
\label{alg:dyn}

Via the bipartitioning procedure, each data point $\vec x_m$ ($m = 1 \ldots n$) is assigned to one out of a finite set of microstates, identified by an index $\mu \in \{1 \ldots N\}$ ($N = 2^b)$. The observed sequence of data points (where the index $m$ enumerates samples taken at consecutive time points) is thereby transformed into a sequence of microstate indices $\mu_m$. Considering this sequence of compound microstates as a realization of a finite-state Markov process, the underlying dynamics is described by a discrete transfer operator $\op P$, an $N \times N$-matrix of transition probabilities between states,
$$
P_{ij} = \prob(\mu_{m + 1} = i ~|~ \mu_m = j),
$$
which may be estimated according to
$$
\hat P_{ij} = \frac{c_{ij}}{\sum_{i'} c_{i'j}}
$$
where
$$
c_{ij} = \#(\mu_{m + 1} = i \land \mu_m = j)
$$
is the number of observed transitions from state $j$ to state $i$.

We assume that the Markov process described by $\op P$ is irreducible, i.e. that it is possible to reach any state from any other state.%
\footnote{We use the terminology and results of \citet{feller:introduction}, Ch.~XV.}
If this is not the case, the system has not only almost invariant but proper invariant sets, each forming an irreducible process of its own which itself may be subjected to a search for almost invariant subsets.%
\footnote{Another possible problem is that there may be states a transition into or out of which is never observed, because they only occur at the beginning or end of the given data segments. Along with the general possibility of transient states, this issue is resolved in a natural way by the reversibilization step described below.}
We assume moreover that the process is aperiodic, which is already the case if only one diagonal element $P_{ii}$ is different from zero. For a finite-state Markov process these two properties amount to ergodicity, which implies that there exists a unique invariant probability distribution $\vec \varpi$ over microstates, with $\op P ~ \vec \varpi = \vec \varpi$, which is also the limit distribution approached from every initial condition.

The analysis of the dynamical properties of a Markov process leading to the identification of its metastable states is strongly facilitated if it is reversible, i.e. if the dynamics is invariant under time reversal: $P_{ij} ~ \varpi_j = P_{ji} ~ \varpi_i$ for all $i, j$. This property cannot usually be assumed for an arbitrary empirically observed process. But since the property of metastability, the tendency of the system to stay within certain regions of the state space for prolonged periods of time, is itself indifferent with respect to the direction of time \citep{froyland:statistically}, we can base the search for the corresponding almost invariant sets on the transition matrix for the reversibilized process $\op R$,
$$
R_{ij} = \frac{1}{2} \left ( P_{ij} + \frac{P_{ji} ~ \varpi_i}{\varpi_j} \right ),
$$
instead. This operator can be directly estimated according to
$$
\hat R_{ij} = \frac{c_{ij} + c_{ji}}{\sum_{i'} (c_{i'j} + c_{ji'})}
$$
i.e. by counting transitions forwards and backwards in time, and the corresponding invariant probability distribution determined as
$$
\hat \varpi_{i} = \frac{\sum_j (c_{ij} + c_{ji})}{\sum_{i'} \sum_j (c_{i'j} + c_{ji'})}.
$$
In the following we will use the symbols $\op R$ and $\vec \varpi$ to denote these estimated quantities.

\subsection{Almost invariant sets}
\label{alg:almost}

To identify almost invariant sets we employ the \textsc{pcca+} algorithm which was developed by \citet{deuflhard:robust} to find metastable states in the conformation dynamics of molecules. In this section we outline the main ideas of that approach which are necessary to understand the operation of the method, while for further details on the implementation and mathematical background the reader is referred to their paper.

Our starting point is the ergodic and reversible Markov process characterized by the $N \times N$-transition matrix $\op R$ along with its invariant probability distribution $\vec \varpi$. Due to the reversibility of the process (symmetry of the stationary flow $R_{ij} ~\varpi_j$), the left and right eigenvectors $\vec A_k$, $\vec p_k$ and eigenvalues $\lambda_k$ of $\op R$,
$$
\vec A_k ~ \op R = \lambda_k ~ \vec A_k,
\quad \op R ~ \vec p_k = \lambda_k ~ \vec p_k,
\quad k = 1\ldots N,
$$
are real-valued. Resolving the scaling ambiguity left by the orthonormality relation $\vec A_k ~ \vec p_l = \delta_{kl}$ by the choice $p_{ik} = \varpi_i ~ A_{ki}$ leads to the normalization equations
$$
\sum_i \frac{p_{ik}^2}{\varpi_i} = 1
\quad \mathrm{and} \quad
\sum_i \varpi_i A_{ki}^2 = 1,
$$
and the transition matrix can be given the spectral representation
$$
\op R = \sum_k \lambda_k ~ \vec p_k \vec A_k.
$$
We assume that the eigenvalues are sorted in descending order, $\lambda_1 \geq \lambda_2 \geq \ldots \geq \lambda_N$. The unique largest eigenvalue $\lambda_1 = 1$ belongs to the invariant probability distribution, $\vec p_1 = \vec \varpi$, while the corresponding left eigenvector has constant coefficients, $A_{1i} = 1$.

If the process $\op R$ possesses $q$ almost invariant sets, it can be seen as the result of a perturbation of a process $\op{\bar R}$ that possesses $q$ perfectly invariant sets. Any (normalized) element of the right eigenvector subspace of $\op{\bar R}$ belonging to eigenvalues $\bar \lambda_1 = \ldots = \bar \lambda_q = 1$ gives an invariant probability distribution of the process. Moreover, if the invariant sets are described by characteristic functions  $\bar \chi_l$ ($l = 1 \ldots q$), such that $\bar \chi_l(i) = 1$ if state $i$ belongs to invariant subset $l$, $0$ otherwise, then any linear combination of them is a left eigenvector of $\op{\bar R}$ for eigenvalue $1$. Conversely, from any linearly independent set of left eigenvectors for eigenvalue $1$, $\{\vec{\bar A}_1, \vec{\bar A}_2, \ldots, \vec{\bar A}_q\}$, the characteristic functions of the invariant sets can be recovered via suitable linear combinations.

Through the perturbation, the multiple eigenvalue $1$ becomes a cluster of large eigenvalues $\lambda_1, \lambda_2, \ldots, \lambda_q$ close to $\lambda_1 = 1$, and the invariant sets become almost invariant sets. They are described by \emph{almost characteristic functions} $\chi_l(i)$, attaining values in the range $[0, 1]$ which may be interpreted as quantifying the degree to which state $i$ belongs to almost invariant set $l$. In analogy to the unperturbed case, these functions are constructed as linear combinations of the left eigenvectors belonging to the $q$ large eigenvalues,
$$
\chi_l(i) = \sum_{k = 1}^q \alpha_{kl} ~ A_{ki},
\quad l = 1 \ldots q,
$$
defined by coefficients $\op \alpha = (\alpha_{kl})$. Admissible are those regular transforms that conform to the constraints
\begin{itemize}
\item partition of unity: $\sum_l  \chi_l(i) = 1$ for all $i$, and
\item non-negativity: $\chi_l(i) \geq 0$ for all $i,l$.
\end{itemize}
The \textsc{pcca+} algorithm optimizes the transform $\op \alpha$ with respect to an objective function to be maximized; we here choose the \emph{maximum scaling} function
$$
I(\op \alpha) = \sum_l \max_i \chi_l(i),
$$
which favors attributions of states $i$ to almost invariant sets $l$ that are as clear-cut as possible.

The input data for the optimization are the dominant left eigenvectors $\vec A_k$, $k = 1 \ldots q$. The first eigenvector is trivially $\vec A_1 = (1, \ldots, 1)$, but the remaining eigenvector coefficients can be geometrically interpreted as attributing to each microstate $i$ a position in a $(q - 1)$-dimensional left \emph{eigenvector space} with position vectors
$$
\vec o(i) = \left ( A_{ki} \right ),
\quad k = 2 \ldots q.
$$
Within this space, the optimization procedure appears as fitting a $q$-simplex as closely as possible around the microstate points. In a system with pronounced metastable macrostates each of them appears as a cluster of microstates located at the boundary of the point cloud, and the optimization procedure matches these $q$ clusters to one of the vertices of the $q$-simplex. The values of the almost characteristic functions $\chi_l(i)$ then attain the geometric meaning of barycentric coordinates of the data points with respect to the locations of the simplex vertices $\vec v_l$:
$$
\vec o(i) = \sum_{l = 1}^q \chi_l(i) ~ \vec v_l.
$$

Finally, metastable macrostates corresponding to almost invariant sets of microstates are identified by attributing each microstate $i$ to that macrostate $l \in \{1 \ldots q\}$ for which the almost characteristic function $\chi_l(i)$ attains the highest value (or, to whose defining vertex it is closest in terms of barycentric coordinates).

\subsection{Macrostates and timescales}
\label{alg:time}

If no prior information on the number of metastable states to be identified is available, it is desirable to obtain an estimate from the data set itself. Since the existence of $q$ almost invariant sets leads to $q$ large eigenvalues, a criterion based on gaps in the eigenvalue spectrum is the natural choice.

However, the concrete values in the spectrum of $\op R$ depend on the step size of the underlying discrete time, which is implicitly given with the input data. Changing the timescale from $1$ to $\tau$ steps, the process has to be described by the transition matrix $\op R^\tau$, whose spectral representation
$$
\op R^\tau = \sum_k \lambda_k^\tau ~ \vec p_k \vec A_k
$$
is essentially the same as that of $\op R$, but with eigenvalues raised to the power $\tau$.

The question which eigenvalues or which gap in the eigenvalue spectrum is to be considered ``large'' therefore depends on the chosen timescale. As \citet{gaveau:dynamical} note, a coarse-graining of the  state space always implies a corresponding ``coarse-graining'' or rather change of scale with respect to the time axis.

We propose%
\footnote{See \citet{allefeld:detecting} for a very similar approach in a different context.}
a measure of the size of spectral gaps that is invariant under a rescaling of the time axis. This is achieved by transforming eigenvalues into associated characteristic timescales,
$$
T(k) = -\frac{1}{\log | \lambda_k |},
$$
and introducing the \emph{timescale separation factor} as the ratio of subsequent timescales:
$$
F(k) = \frac{T(k)}{T(k + 1)} = \frac{\log | \lambda_{k+1} |}{\log | \lambda_k |}.
$$
Substituting $\lambda_k^\tau$ for $\lambda_k$ in this equation, the resulting factors cancel out, so that $F(k)$ provides a measure of the spectral gap between eigenvalues $\lambda_k$ and $\lambda_{k+1}$ that is independent of the timescale.

Using this measure, the number of macrostates $q$ is estimated as the value of $k$ for which $F(k)$ becomes maximal. The choice $q = 1$ leading to a single macrostate comprising all microstates has thereby to be excluded, because it is always associated with the largest timescale separation factor, $F(1) \rightarrow \infty$.

If several larger gaps exist, a ranking list of possible $q$-values may be compiled, where each value leads to a different possible coarse-graining of the system into macrostates. This way different layers of the system's dynamical structure are recovered, which (extending \citeauthor{deuflhard:robust}'s approach) may be considered as the result of multiple superimposed perturbations.  An example of this is given in the following section, where the method is illustrated using data from a simulated system.

\section{Example: A system with four metastable macrostates}
\label{example}

To illustrate the operation of the algorithm we apply it to data from a simulated system, where we can interpret the analysis results with respect to our precise knowledge of the underlying dynamics. We use a discrete-time stochastic system in two dimensions, $(x_1, x_2)$, where the change over each timestep is given by
$$
\Delta x_i = a \left ( x_i - 2 x_i ^ 3 \right ) + b_i \xi_i,
$$
with $a = 0.01$, $(\xi_1, \xi_2)$ standard normal two-dimensional white noise, $b_1 = 0.03$, and $b_2 = 0.05$. The first term of the right hand side of this equation describes an overdamped movement within a double-well potential along each dimension, leading to four attracting fixed points at $(x_1, x_2) = (\pm 1/\sqrt{2}, \pm 1/\sqrt{2})$. Without the stochastic second term, the system would be decomposable into four invariant sets, separated by the two coordinate axes. But due to the noise the system performs a random walk, staying for prolonged periods of time in the vicinity of one of the attracting points, but occasionally wandering into another point's basin of attraction. These switches occur more frequently along $x_2$ because the noise amplitude is larger in that direction, $b_2 > b_1$.

\begin{figure}
\centering \includegraphics{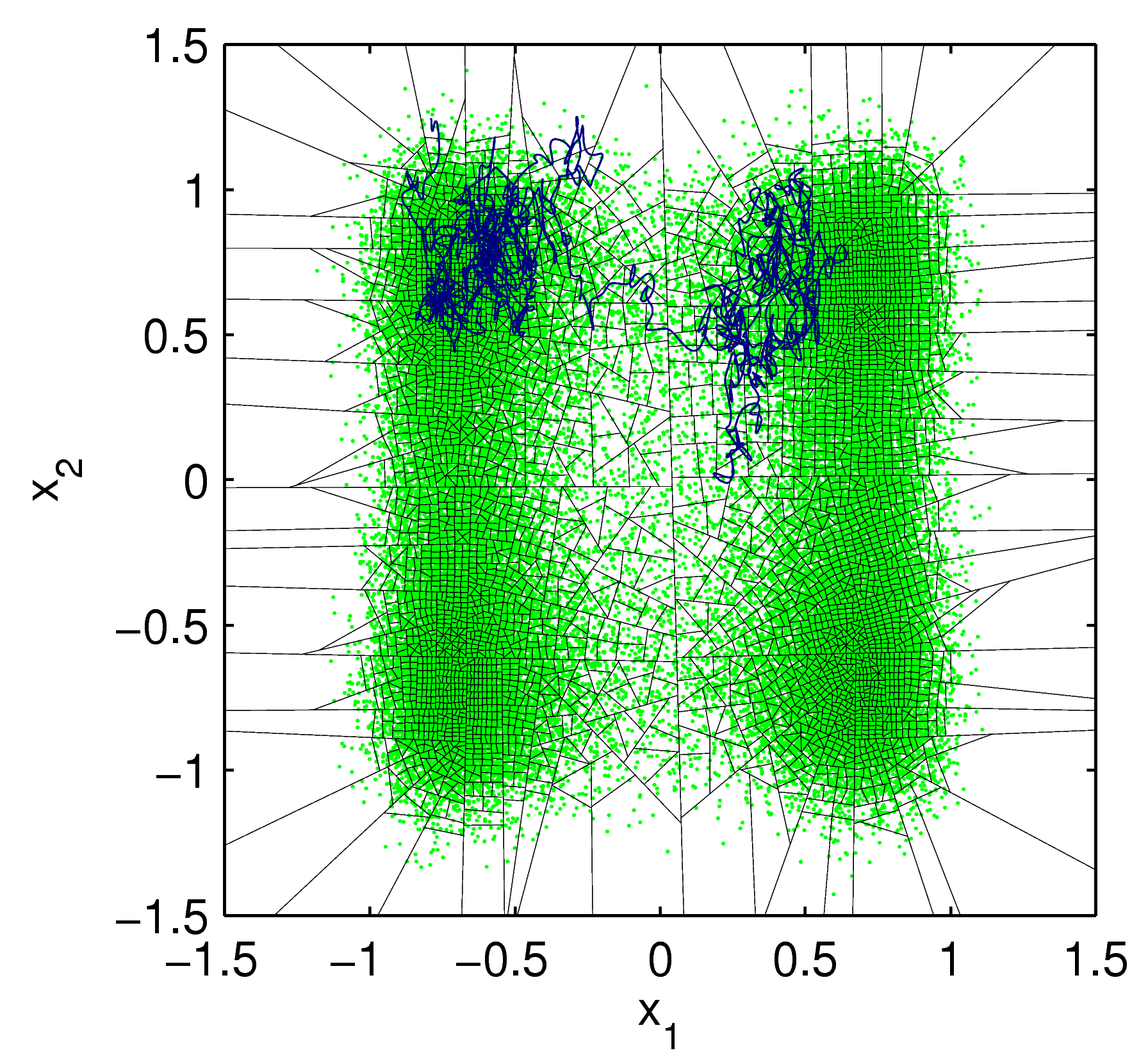}
\caption{Data points from a simulation run of a system with four metastable macrostates over $10^6$ time steps, and a part of the connecting trajectory. Straight lines indicate the cell borders of the partition into $4096$ microstates obtained via the bipartition algorithm.}
\label{4phase_a}
\end{figure}

\begin{figure}
\centering \includegraphics{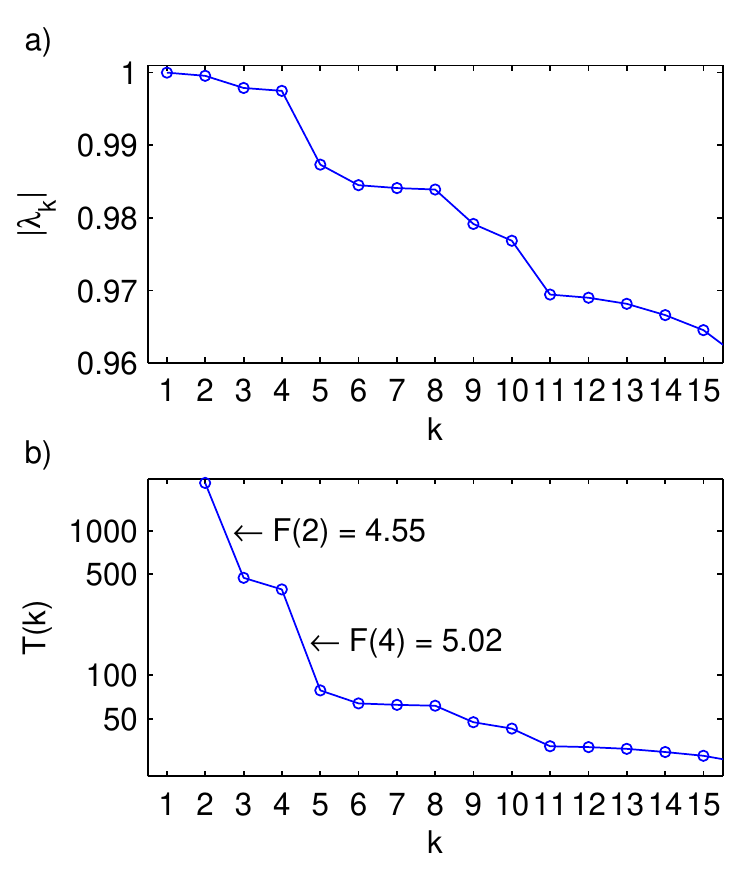}
\caption{Eigenvalue spectrum of the transition matrix $\op R$ of the system with four metastable macrostates. (a)~The eigenvalues of largest magnitude. (b)~Logarithmic timescales; locations and values of the two largest timescale separation factors are indicated.}
\label{4phase_b}
\end{figure}

\begin{figure}
\centering \includegraphics{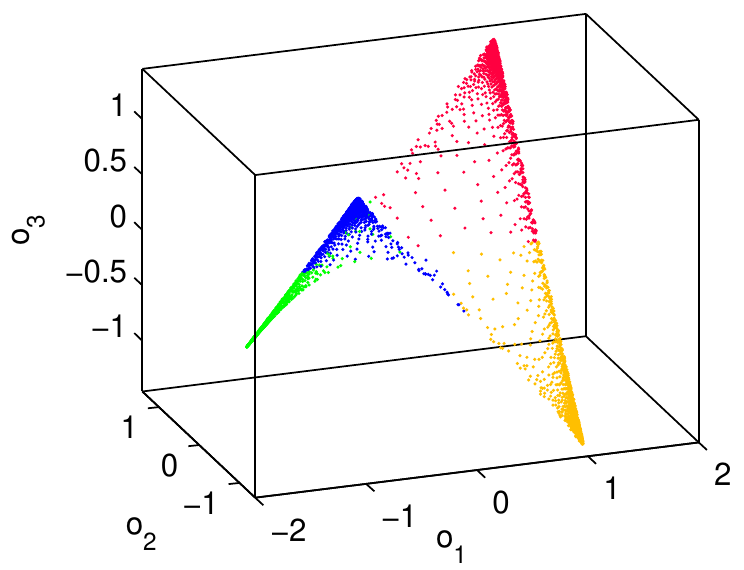}
\caption{Eigenvector space $(o_1, o_2, o_3)$ of the system with four metastable macrostates for $q = 4$. Each dot representing a microstate is colored according to which vertex of the enclosing tetrahedron is closest, defining the four metastable macrostates.}
\label{4phase_c}
\end{figure}

\begin{figure}
\centering \includegraphics{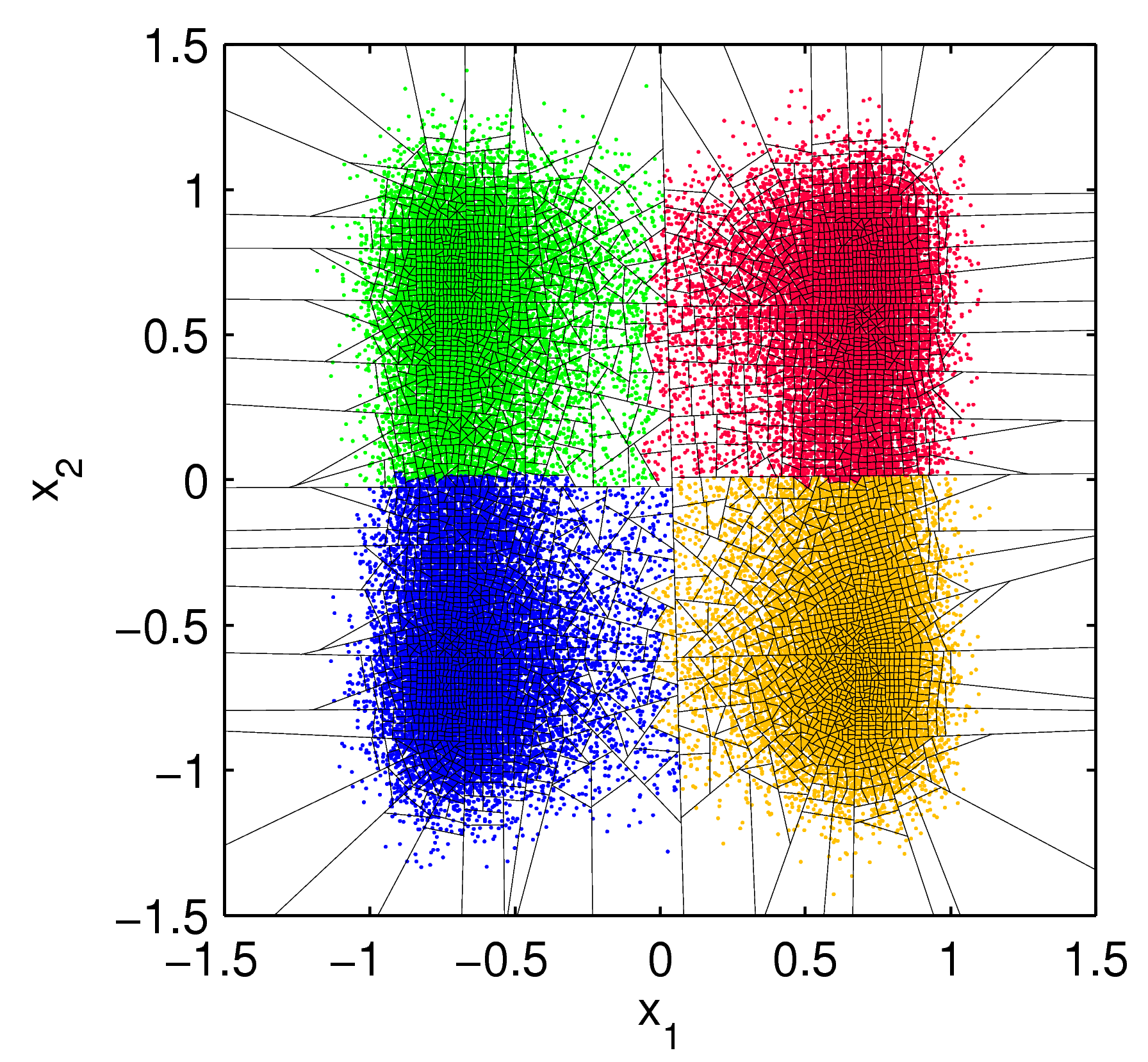}
\caption{Micro- and macrostates of the system with four metastable macrostates. Lines indicate the cell borders of the partition of the state space into microstates, while the coloring of data points shows the attribution of microstates to one of the four metastable macrostates.}
\label{4phase_d}
\end{figure}

Data resulting from a simulation run of this system are shown in Fig.~\ref{4phase_a}. A section of the connecting trajectory illustrates how the system state moves through the state space, entering and leaving the cells of the microstate partition. Counting these transitions between cells leads to an estimate of the reversibilized transition matrix $\op R$.

The largest eigenvalues of $\op R$ are plotted in Fig.~\ref{4phase_b}a, revealing a group of four large eigenvalues ($> 0.995$), which itself is subdivided into two groups of two eigenvalues each. This picture becomes clearer after the transformation into timescales $T(k)$. In Fig.~\ref{4phase_b}b they are displayed on a logarithmic scale, such that the magnitude of timescale separation factors $F(k)$ becomes directly visible in the vertical distances between subsequent data points. The largest separation factor is $F(4) = 5.02$, closely followed by $F(2) = 4.55$, indicating that a partitioning of the state space into $q = 4$ macrostates is optimal, while searching for two different macrostates may also yield a meaningful result.

The identification of almost invariant sets of microstates defining the metastable macrostates is performed within the 3-dimensional eigenvector space $(o_1, o_2, o_3)$. Fig.~\ref{4phase_c} reveals that the points representing microstates are located on a saddle-shaped surface  stretched out within a 4-simplex or tetrahedron. The algorithm identifies the vertices of the tetrahedron and attributes each microstate to that macrostate whose defining vertex is closest, resulting in the depicted separation into four sets.

In Fig.~\ref{4phase_d} this result is re-translated into the original state space of Fig.~\ref{4phase_a}, by coloring the data points of each microstate according to the macrostate it is assigned to. The identified metastable states coincide roughly with the basins of attraction of the four attracting points, i.e., the almost invariant sets of the system's dynamics.

From Fig.~\ref{4phase_c} we can also assess which macrostate definitions would be obtained by choosing $q = 2$, the next-best choice for the number of metastable states according to the timescale separation factor criterion. In this case the eigenvector space is spanned by the single dimension $o_1$, along which the two vertices of the tetrahedron on the left and right side, respectively, coincide. This means that the two resulting macrostates each consist of the union of two of the macrostates obtained for $q = 4$. With respect to the state space, these two macrostates correspond approximately to the areas $x_1 > 0$ and $x_1 < 0$.

This result can be understood from the system's dynamics, since because of the smaller probability of transitions along $x_1$ these two areas of the state space form almost invariant sets, too. As can be seen from this example, the possibility to select different $q$-values of comparably good rating may allow to recover different dynamical levels of a system, giving rise to a hierarchical structure of potential macrostate definitions.

\section{Application to EEG data}
\label{application}

For the purpose of a first application of the algorithm to neurophysiological data, we chose an electroencephalographic (EEG) recording from a patient suffering from petit-mal epilepsy, a condition characterized by the occurrence of frequent short (several seconds) epileptic episodes, during which the patient becomes irresponsive \citep[cf.][]{niedermeyer:abnormal}. This kind of data is favorable for our methodological approach because we can expect two clearly distinct states to be present---``normal'' EEG / mentally present and paroxysmal episodes / mentally absent---, and because it is possible to observe many transitions between these states in a recording of moderate size.

The data set consists of a section of 89\,min length from the patient's monitoring EEG. It was recorded from the 19 electrode positions of the international 10-20 system \citep{aes:guidelines} at a sampling rate of 250\,Hz, digitally bandpass-filtered (2--15\,Hz), and transformed to the average reference. Due to artifact removal by visual inspection the amount of data available for analysis was reduced to 71\,min total length (1\,064\,435 data points).

In the preceding simulation example we know by definition that the given values of the system variables immediately specify its dynamical state, and therefore can be directly processed by the algorithm for the identification of metastable states. With measurement data like EEG the situation is not so clear. The data ``as is'' may be accepted as a specification of the system state, but any further processed version of them fulfills this function as well and may for some reason be even more suitable. This means that empirical data pose the problem of how to define the input state space for the analysis.

For low-dimensional nonlinear deterministic dynamical systems techniques have been developed to reconstruct the state space of the system, or a higher-dimensional space comprising it, from scalar time series via the method of time-delay embedding \citep{takens:detecting, kantz:nonlinear}. However, these techniques are not appropriate for our purposes. Firstly, the data-set is already multi-dimensional and using the embedding approach we would have to either blow up the dimensionality even more, thereby introducing a high amount of redundance, or discard many of the input data channels, possibly loosing crucial information. And secondly, previous attempts to demonstrate low-dimensional nonlinear structure in EEG data had only limited success \citep[cf.][]{theiler:reexamination, palus:nonlinearity}.

Instead, we pursue the following strategy: In a first step, we use the original 19-dimensional data space as the input state space. Guided by the results obtained in this way as well as by independent observations on the behavior of multichannel EEG, in a second step we develop a preprocessing procedure defining a more abstract input state space.

\subsection{Original data state space}

\begin{figure}
\centering \includegraphics{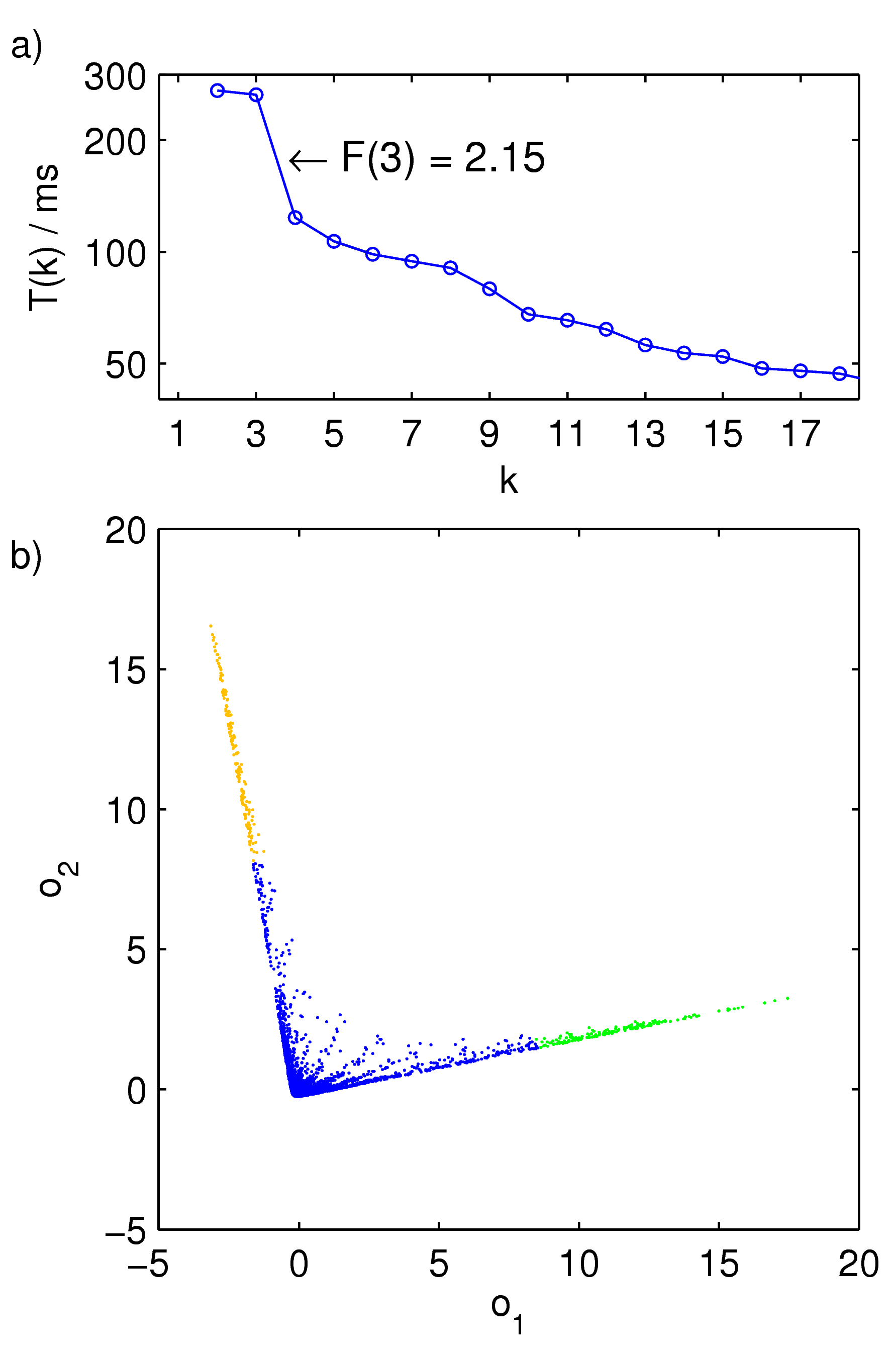}
\caption{Analysis results for the original EEG data state space. a)~Timescale spectrum; a large separation factor indicates three metastable states. b)~Microstate positions in two-dimensional eigenvector space forming a triangular structure, and the resulting three metastable macrostates.}
\label{original_eigen}
\end{figure}

Using the recursive bipartitioning algorithm, the data points were assigned to 32\,768 different compound microstates (32 or 33 points in each cell). The resulting timescale spectrum (Fig.~\ref{original_eigen}a) exhibiting a large separation factor $F(3) = 2.15$ suggests a search for three metastable macrostates in a two-dimensional eigenvector space (Fig.~\ref{original_eigen}b). This is supported by the 3-simplex shape of the distribution of microstate positions within this space.

\begin{figure}
\centering \includegraphics{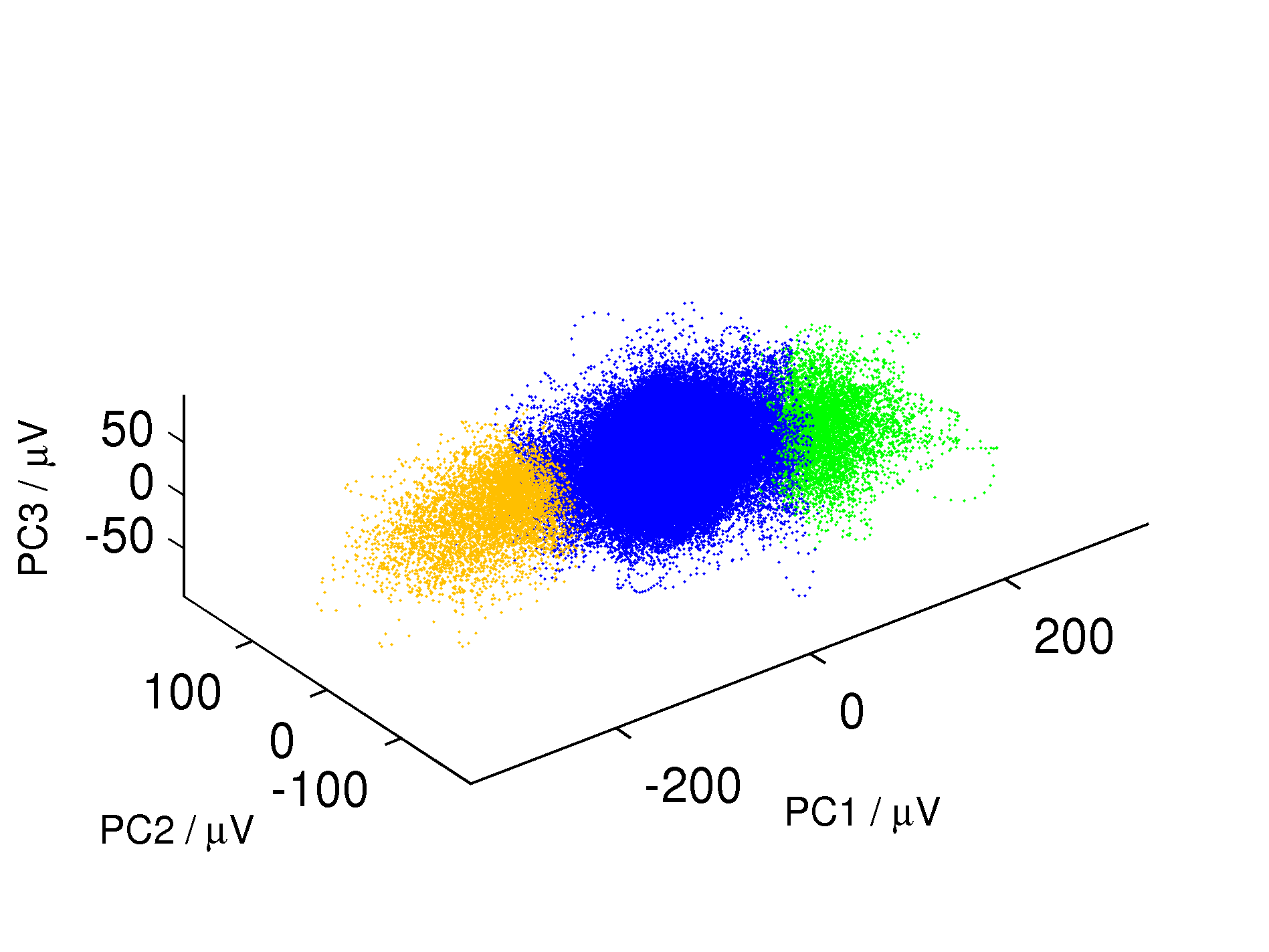}
\caption{Analysis results for the original EEG data state space: Location of data points belonging to the three identified metastable states. The 19-dimensional state space is represented using the first three PCA components of the data.}
\label{original_pc}
\end{figure}

The arrangement of the areas belonging to the identified macrostates in the input data space is shown in Fig.~\ref{original_pc}, where the 19-dimensional space is represented using the first three PCA components of the data. The most prevalent state accounting for about 99\% of the data points appears here as a centrally located spherical area, with the two other states forming handle-like appendices at opposite sides.

\begin{figure*}
\centering \includegraphics{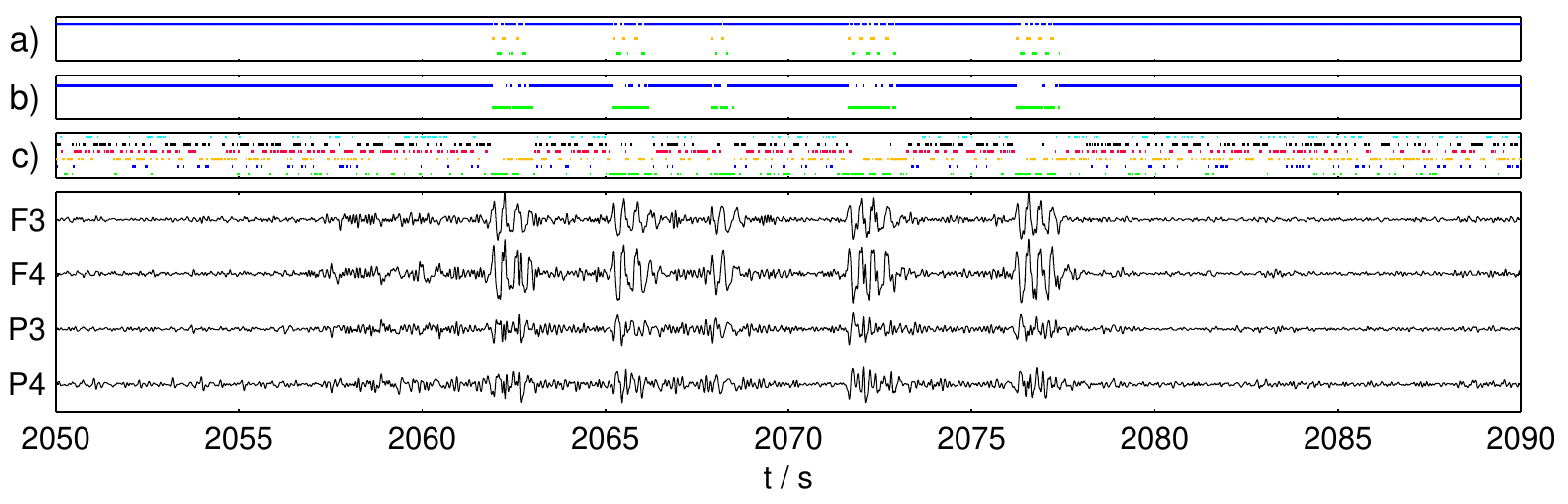}
\caption{Analysis results illustrated using a segment of 40\,s length. Upper panels: Macrostate dynamics over time, resulting from different input state space definitions. a)~Original EEG data state space. b)~Amplitude vector state space. c)~As in (b), but using normalized amplitude vectors. Lower panel: EEG timeseries at four selected recording sites. Paroxysmal episodes are characterized by short bursts of spike-wave activity.}
\label{eeg}
\end{figure*}

The role of these three macrostates becomes clearer considering the transitions between them over time, in comparison with the underlying EEG time series (Fig.~\ref{eeg}a). Within periods of normal electroencephalographic activity the system stays within the ``main'' macrostate, while during seizures switches between all three states occur regularly, corresponding to an oscillation along the PC1 axis of Fig.~\ref{original_pc}. This macrostate dynamics reflects the spike-wave oscillatory activity visible in the EEG channels shown in the lower panel of Fig.~\ref{eeg}, which are characteristic for paroxysmal episodes.

With this first result, the attempt at identifying emergent macrostates using the EEG data space is only partially successful: The occurrence of states correlates strongly with those features of the underlying process which are psychophysiologically most important, and also most prominent in visual inspection of the data. However, the two states expected are not directly recovered by the EEG analysis. Instead of one persistent state during paroxysmal episodes, we find rapid oscillatory changes between states including the one associated with normal EEG. This indicates that the input state space is not yet optimally defined.

\subsection{Amplitude vector state space}

This finding can be understood from the fact that electroencephalographic activity in general is so strongly shaped by a predominant oscillatory layer of the dynamics---not only during epileptic episodes but also in normal EEG, particularly in the form of the alpha rhythm---that it is hard to discern more subtle dynamical features. To recover those features, we need a preprocessing step that eliminates the oscillatory character of the data but retains the more slowly changing parameters of the oscillation.

\begin{figure}
\centering \includegraphics{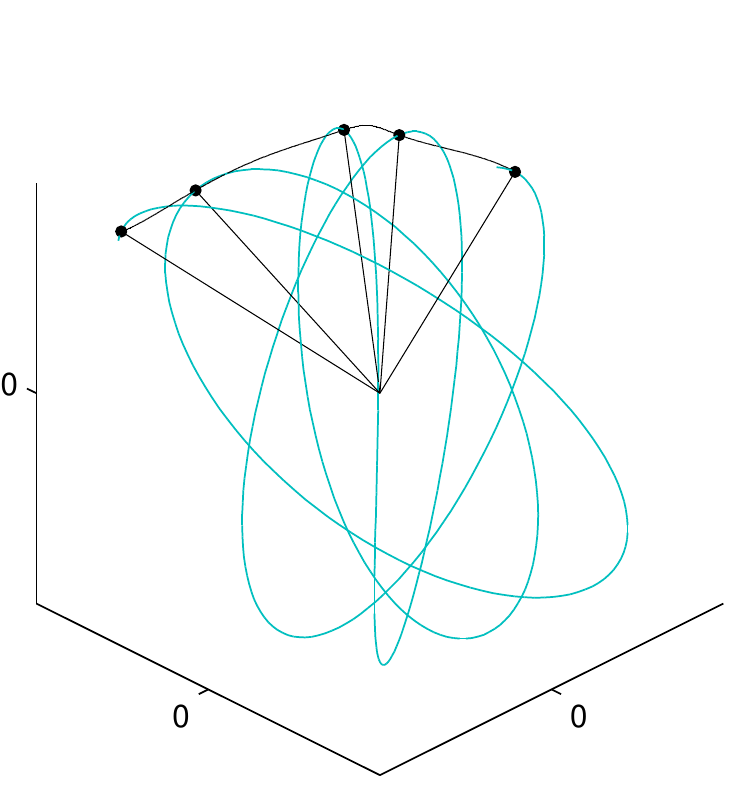}
\caption{Amplitude vector state space. The trajectory corresponding to a multivariate oscillatory signal like EEG takes on the form of an elliptical orbit with slowly varying parameters. Locally matching ellipses to the trajectory, the instantaneously dominant oscillatory component can be characterized by the major semiaxis vectors (straight radial lines), resulting in a description of the system's oscillatory state which itself evolves in a non-oscillatory way (black curve).}
\label{ellipse}
\end{figure}

As observed by \citet{wackermann:segmentation}, the trajectory formed by multichannel EEG within the data space can be approximated by a movement along an elliptical orbit with slowly changing orientation and shape (Fig.~\ref{ellipse}). By locally matching ellipses to the data, a global instantaneous phase and amplitude can be defined, where the amplitude corresponds to the two main semiaxis vectors of the ellipse. (For a full account of the calculation see App.~\ref{appendix_instantaneous}.) For simplicity we only use the major semiaxis vector, which specifies the direction and strength of the momentarily dominant oscillatory component, to define an amplitude vector state space as the input state space for the algorithm.%
\footnote{This approach is similar to one of the strategies employed in the ``spatial analysis'' of EEG \citep{lehmann:principles}, to select only those EEG potential maps (data vectors) which occur at local maxima of the ``global field strength'' (the norm of the data vectors).}

\begin{figure}
\centering \includegraphics{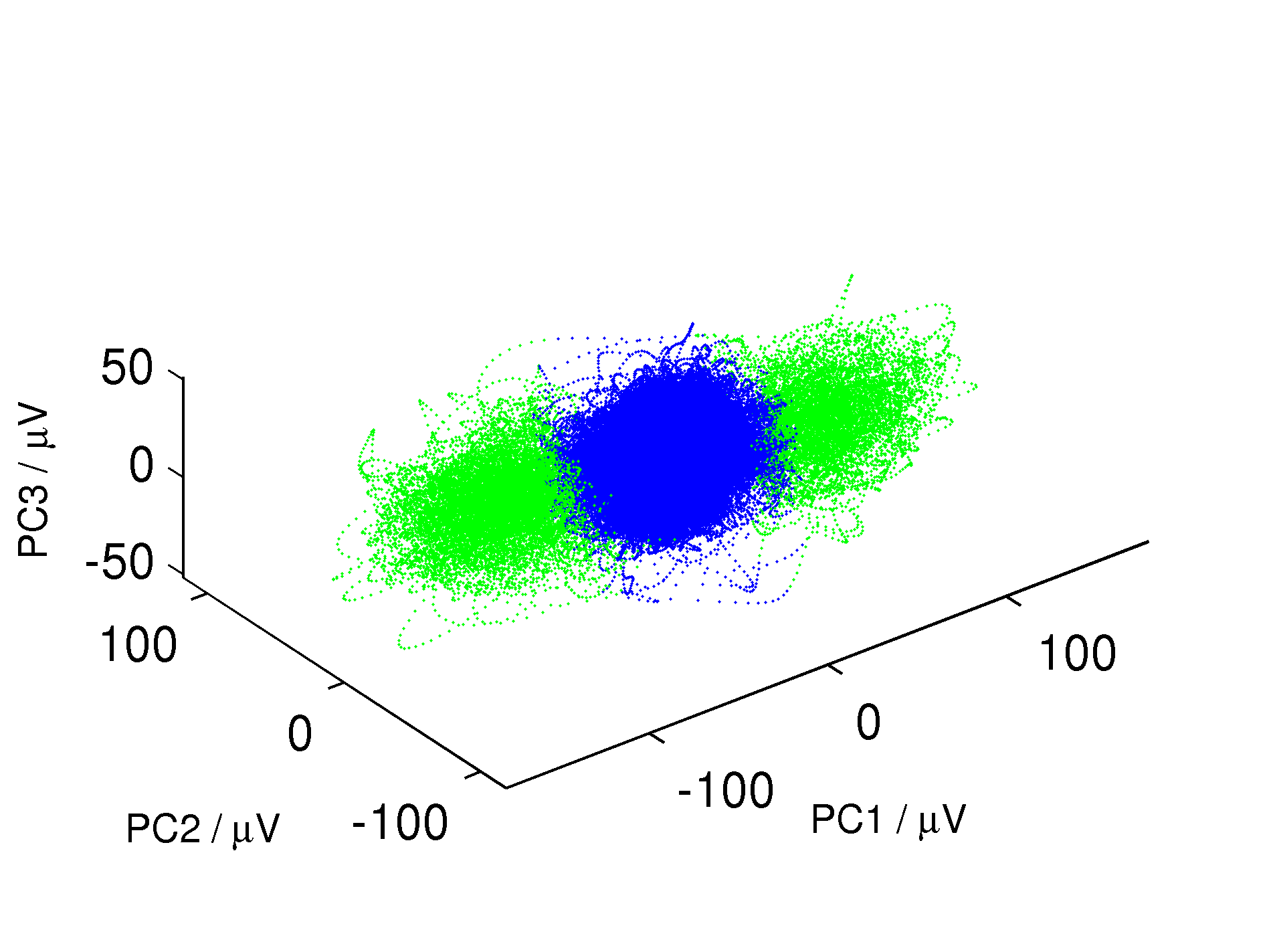}
\caption{Analysis results for the amplitude vector state space: Location of data points belonging to the two identified metastable states in a representation of the state space using the first three PCA components.}
\label{amplitude_pc}
\end{figure}

With the specification of the system state via the major amplitude vector an ambiguity arises, because vectors of opposite orientation are equivalent. This is resolved by enforcing positive sign for the first vector component during the assignment of data points to microstates. For visualization (Fig.~\ref{amplitude_pc}) the axis vectors are used as they come out of the calculation described in App.~\ref{appendix_instantaneous}, that is with basically random orientation.

\begin{figure}
\centering \includegraphics{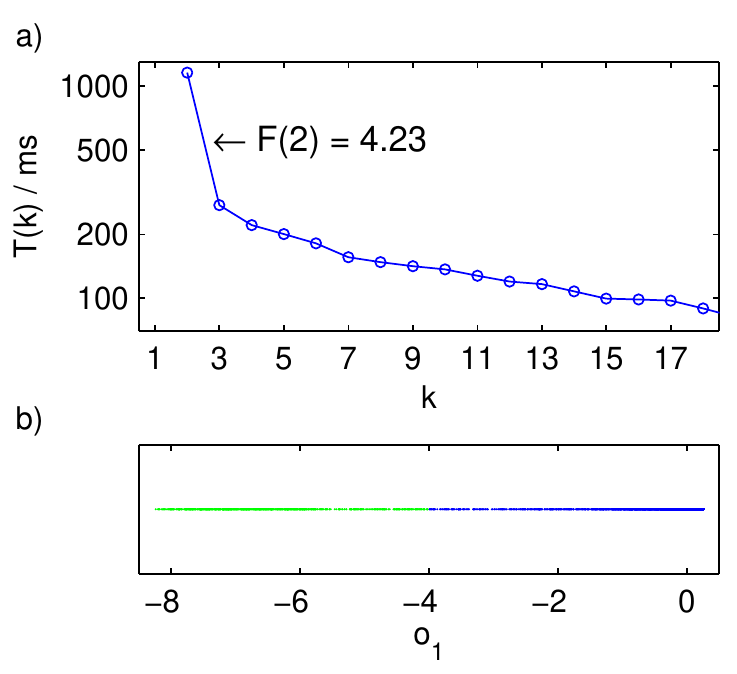}
\caption{Analysis results for the amplitude vector state space. a)~Timescale spectrum indicating the presence of two metastable states. b)~Microstate positions in one-dimensional eigenvector space and the resulting two metastable macrostates.}
\label{amplitude_eigen}
\end{figure}

The resulting timescale spectrum is shown in Fig.~\ref{amplitude_eigen}a. The largest separation factor of $F(2) = 4.23$ now gives a more definite indication of the number of macrostates than for the original data state space. In the corresponding one-dimensional eigenvector space (Fig.~\ref{amplitude_eigen}b) the two macrostates are trivially defined by a cut at the center of the range of values.

In Fig.~\ref{amplitude_pc}, the location of data points belonging to the two macrostates is shown using the first three PCA components of the data points in the amplitude vector state space. Again, a prevalent macrostate (accounting for 98\% of the data points) fills a spherically shaped central area, while two appendices protruding on opposite sites together constitute the second macrostate. Despite the fact that the overall shape of the data cloud is similar to that shown in Fig.~\ref{original_pc}, the reader should keep in mind that the two diagrams depict differently defined state spaces represented with respect to a different set of dimensions.

Fig.~\ref{eeg}b demonstrates that the revision of the input state definition successfully eliminates the oscillatory switching between states during paroxysmal episodes. Starting from the amplitude vector input state space, the algorithm for the identification of metastable states is able to consistently associate normal EEG with one macrostate, and---except for short relapses---epileptic EEG with another macrostate.

\begin{figure}
\centering \includegraphics{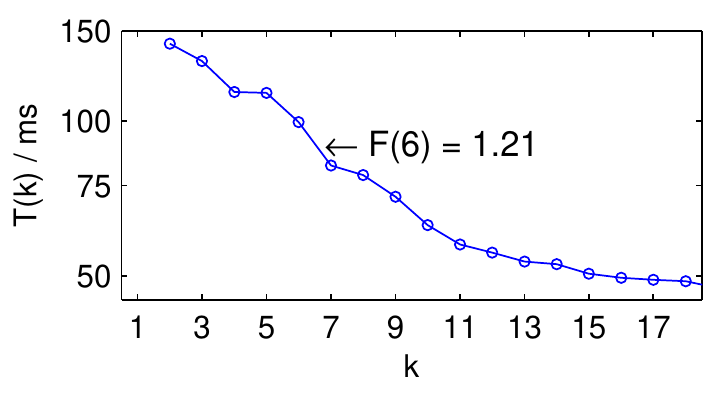}
\caption{Timescale spectrum obtained using normalized axis vectors. The largest separation factor (for six macrostates) is only marginally larger than the other occurring values, indicating that no adequate definition of metastable states is possible.}
\label{normalized_eigen}
\end{figure}

The macrostate structure of the amplitude vector state space shown in Fig.~\ref{amplitude_pc} suggests that the distinction of the two macrostates relies only on the length of the amplitude vector. To check this, we tested the performance of the algorithm when normalized amplitude vectors are used. The timescale spectrum (Fig.~\ref{normalized_eigen}), with a maximal separation factor $F(6) = 1.21$ not substantially larger than the rest, indicates that the identification of macrostates is severely impaired under these circumstances. Even so, an examination of the state dynamics over time (Fig.~\ref{eeg}c) reveals that there are still two states that are mainly attained during epileptic episodes.

\section{Conclusion}

Relations between mental (psychological) and neural (physiological) phenomena form the generally accepted basis for work in various disciplines such as psychiatry, psychophysiology, and cognitive neuroscience. While a large body of knowledge has been gathered in these fields, the conceptual question of how mind and brain are related in precise terms is still largely unresolved. Starting from the notion of the mental as ``emerging'' from neural processes, we argue that this relation of emergence should be understood as one between different descriptions of the same system.

Utilizing concepts from the theory of dynamical systems for the formulation of descriptions, we propose that the relation between descriptive levels should take on the form of a partition or coarse-graining of the state space that is characterized by a preservation of the Markov property. To empirically test the validity of our approach, we turn to a form of such a Markov coarse-grainining which can be algorithmically obtained: that of metastable states. We describe how metastable macrostates of a dynamics observed in empirical data can be identified based on the spectral analysis of the transition matrix governing the microstate dynamics, and illustrate its operation with simulation data.

We apply the method to a recording of electroencephalographic (EEG) data from a human subject suffering from petit-mal epilepsy. Combined with a suitable preprocessing procedure, the algorithm is able to automatically identify metastable states from the data which closely correspond to the mental states of the subject (mentally present / absent). This first application substantiates the practical viability of our approach and appears promising for the future application of the method to more challenging forms of data.

Finally we want to point out that the concept of metastable macrostates in the application to EEG data is similar to the notion of ``brain functional microstates'' introduced by Lehmann and co-workers, which are defined as brief periods of time during which the spatial distribution of the brain's electrical field remains relatively stable \citep{lehmann:multichannel, lehmann:eeg}. Transitions between such states are characterized by an abrupt change of the field topography, allowing to decompose the stream of EEG data into segments of the order of magnitude of 10--100\,ms duration which can usually be grouped into a small number ($< 10$) of classes. Note however that the ``microstate analysis'' of Lehmann et al. results in a coarse-grained description of the brain's electrical activity, i.e. in our nomenclature, a definition of macrostates.

\section*{Acknowledgments}

The authors would like to thank P.~beim Graben for discussions, G.~Froyland for helpful hints, and V.~Krajča and S.\,E.~Petránek (University Hospital Na Bulovce, Prague, Czech Republic) for providing us with the EEG recording.

\appendix
\section{Instantaneous amplitude and phase for multivariate timeseries}
\label{appendix_instantaneous}

The local oscillatory behavior of a real-valued univariate signal $x(t)$ is commonly characterized using the corresponding complex-valued analytic signal $z(t)$ \citep{gabor:theory}. It is obtained by combining $x(t)$ with an imaginary part,
$$
z(t) = x(t) + \ii ~ y(t),
$$
which is defined as the Hilbert transform of $x$,
$$
y(t) = \hilbert x(t) = \frac{1}{\lpi} \, \textrm{\scriptsize P.V.} \!\!\! \int_{-\infty}^\infty \frac{x(t')}{t - t'} \dd t',
$$
where P.V. denotes the Cauchy principal value of the integral. Under the condition that $x(t)$ is dominated by a single frequency component, its instantaneous amplitude $A(t)$ and phase $\phi(t)$ can be determined via the analytic signal according to
$$
A(t) = |z(t)|,
\quad 
\phi(t) = \arg z(t),
$$
so that
$$
x(t) = A(t) ~ \cos \phi(t)
$$
or
$$
z(t) = A(t) ~ \exp(\ii ~ \phi(t)).
$$
The terms amplitude $A$ and phase $\phi$ as they are used here can be interpreted such that they specify the parameters of a strictly periodic sinusoidal oscillation which \emph{locally matches} the behavior of the observed signal $x(t)$ at a given instant $t$. In particular, $\phi(t)$ attains the value 0 (or equivalently, an integer multiple of $2 \lpi$) whenever the actual value of $x(t)$ coincides with the associated instantaneous amplitude $A(t)$.

These properties of the analytic signal can also be utilized to determine the parameters of the locally matching oscillation for a multivariate signal $\vec x(t) = \left ( x_i(t) \right )$ ($i = 1 \ldots K$). We assume that each component signal $x_i(t)$ is dominated by a single frequency and that the frequencies of different signals are similar. Using $\vec y(t)$ to denote the channel-wise Hilbert transform of $\vec x(t)$ and $\vec z(t)$ for its channel-wise completion to the analytic signal, the local extension of the signal's oscillatory behavior for instant $t$ is obtained with
$$
\vec z_t(\theta) = \vec z(t) ~ \exp(\ii ~ \theta),
$$
parametrized by $\theta \in [0, 2 \lpi]$. Its real part
$$
\vec x_t(\theta) = \vec x(t) \cos \theta - \vec y(t) \sin \theta
$$
gives the multivariate oscillation that locally matches the behavior of the signal at instant $t$; its trajectory is an \emph{elliptical orbit} with conjugate axes specified by the vectors $\vec x(t)$ and $\vec y(t)$.

\begin{figure}
\centering \includegraphics{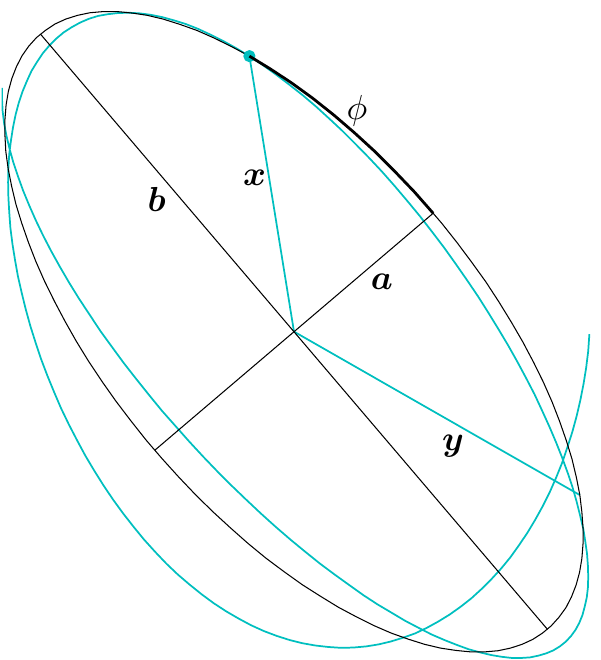}
\caption{Determination of local ellipse axes. The trajectory formed by the multivariate signal is locally matched to an elliptical orbit, which is defined by the data vector $\vec x$ at a given instant and the corresponding vector $\vec y$ from the signal's channel-wise Hilbert transform as conjugate semiaxis vectors. Main semiaxis vectors of the ellipse, $\vec a$ and $\vec b$, are obtained using the associated multivariate instantaneous phase $\phi$.}
\label{ellipse_app}
\end{figure}

From these conjugate axes, the main semiaxis vectors $\vec a(t)$ and $\vec b(t)$ of the local ellipse can be calculated (Fig.~\ref{ellipse_app}). It proves useful to do so via introducing a global (channel-independent) instantaneous phase $\phi(t)$, such that for $\phi(t)$ $\in$ $ \{ 0, \tfrac{1}{2}\lpi,$ $\lpi, \tfrac{3}{2}\lpi \}$ or equivalents, $\vec x(t)$ coincides with one of the main semiaxis vectors or its negative. This is achieved choosing
$$
\phi(t) = \frac{1}{2} \arctan \frac{2 ~ \vec x(t) \cdot \vec y(t)}{|\vec x(t)|^2 - |\vec y(t)|^2}.
$$
Since the resulting values in the range $[-\frac{\lpi}{4}, \frac{\lpi}{4}]$ cover only one quarter of a cycle, the outcome may be transformed into an equivalent but more useful representation via a standard ``unwrapping'' procedure (adding or subtracting $\frac{\lpi}{2}$ at discontinuity points) to enforce a smooth evolution of $\phi(t)$.

Using this result, main semiaxis vectors of the locally matching ellipse at instant $t$ are obtained by going backwards along $\vec x_t(\theta)$ by an amount of $\phi(t)$ or forwards by $\frac{\lpi}{2} - \phi(t)$, i.e.
$$
\vec a(t) = \vec x_t \left ( -\phi(t) \right )
$$
and
$$
\vec b(t) = \vec x_t \left ( \frac{\lpi}{2} - \phi(t) \right ) .
$$
If $\phi(t)$ has been adjusted for a smooth evolution over time, the same can be expected from the resulting $\vec a(t)$ and $\vec b(t)$. It is, however, not clear from this definition which one of these vectors specifies the major and minor axis of the ellipse, respectively, and it is possible that over the course of time the two vectors change roles. For a specific application of this result, further processing may therefore be necessary.

Complementary to the generalization of the instantaneous phase concept, a multivariate instantaneous amplitude $\vec A(t)$ can be defined such that
$$
\vec z(t) = \vec A(t) ~ \exp (\ii ~ \phi(t)),
$$
which is given by
$$
\vec A(t) = \vec a(t) - \ii ~ \vec b(t).
$$
The channel-wise modulus of this quantity corresponds to the instantaneous amplitudes $A_i(t)$ of the component signals, while the argument comprises the phase differences between the global and the component signal instantaneous phases, $\phi_i(t) - \phi(t)$.

\bibliography{emergent}

\end{document}